\documentclass[a4paper,11pt]{article}
\pdfoutput=1 

\usepackage{jcappub} 

\usepackage[T1]{fontenc} 
\usepackage{latexsym}
\usepackage{graphics}
\usepackage{graphicx}
\usepackage{dcolumn}
\usepackage{bm}
\usepackage{natbib}
\usepackage{amsfonts}
\usepackage{amsmath}
\usepackage{multirow}
\usepackage{aas_macros}

\title{\boldmath Hydrostatic equilibrium configurations of neutron stars in the $f(R,\mathcal{L},T)$ gravity theory}

\author[a, 1]{J. A. S. Fortunato,\note{Corresponding author.}}
\author[b]{P.H.R.S. Moraes,}
\author[c]{E. Brito,}
\author[d]{J.G. de Lima J\'unior,}
\author[e]{T.S. Guerini}

\affiliation[a]{PPGCosmo, CCE, Universidade Federal do Espírito Santo (UFES), Av. Fernando Ferrari, 540, CEP 29.075-910, Vitória, ES, Brazil}
\affiliation[b]{Laborat\'orio de F\'isica Te\'orica e Computacional (LFTC),
 Universidade Cidade de S\~ao Paulo (UNICID) - Rua Galv\~ao Bueno 868, 01506-000 S\~ao Paulo, SP, Brazil}
\affiliation[c]{Centro de Ci\^encias Exatas e das Tecnologias, Universidade Federal do Oeste da Bahia - Rua Bertioga 892, 47810-059, Barreiras, BA, Brazil}
\affiliation[d]{Instituto de F\'isica, Universidade Federal do Alagoas (UFAL) - Avenida Lourival Melo Mota S/N, 57072-970, Macei\'o, AL, Brazil}
\affiliation[e]{Instituto de F\'isica, Universidade do Estado do Rio de Janeiro (UERJ) - Rua S\~ao Francisco Xavier 524, 20550-013 Rio de Janeiro, RJ, Brazil}

\emailAdd{jeferson.fortunato@edu.ufes.br}
\emailAdd{moraes.phrs@gmail.com}
\emailAdd{eliasbaj@ufob.edu.br}
\emailAdd{grimario.lima@fis.ufal.br}
\emailAdd{thais.silvaguerini@gmail.com}

\abstract{In the present work, we obtain the hydrostatic equilibrium configurations of neutron stars in the recently proposed $f(R,\mathcal{L},T)$ theory of gravity, for which $R$ is the Ricci scalar, $\mathcal{L}$ is the matter lagrangian density, $T$ is the trace of the energy-momentum tensor and $f$ is a function of the argument. This theory emerges in the present literature as a generalized geometry-matter coupling theory of gravity. We derive the Tolman-Oppenheimer-Volkoff-like equation for a particular functional form of the $f(R,\mathcal{L},T)$ function. Our solutions are obtained from realistic equations of state describing matter inside neutron stars. We obtain stable solutions for neutron stars and we show that for some values of the free parameter of the theory it is possible to be in agreement with both NICER and LIGO/Virgo observational data.}

\begin{document}
\maketitle
\flushbottom
\section{Introduction}\label{sec:int}

The astrophysics of compact objects is a very engaged field of Physics nowadays. The recent first detection of gravitational waves has certainly strongly contributed to it \cite{abbott/2016}. The gravitational wave sign reported in \cite{abbott/2016} was generated by two coalescing black holes of $36^{+5}_{-4}$M$_\odot$ and $29\pm4$M$_\odot$ at redshift $z=0.09^{+0.03}_{-0.04}$. Later, the first detection of a gravitational wave sign coming from a coalescing neutron star binary system happened \cite{abbott2017}. Such an event also gave birth to the ``Multi-Messenger Era'' of Astrophysics \cite{abbott/2017b}, as it was the first astrophysical event to be detected in both electromagnetic and gravitational wave spectra.  

A remarkable compact object was reported in the GW190814 gravitational wave event  \cite{abbott/2020}. This event is characterized by a $22.2-24.3$M$_\odot$ black hole and a neutron star with $2.50-2.67$M$_\odot$. Although its secondary object could be the lighest black hole ever detected, most of the current literature points to the object as a massive neutron star \cite{huang/2020,roupas/2021,tsokaros/2020,wu/2021,fattoyev/2020,dexheimer/2021}.

To explain such a massive neutron star has been challenging for theoretical physicists. As we still do not know the neutron star equation of state, a huge number of models has been proposed in the literature, as one can check, for instance, \cite{baym/2019,somasundaram/2023,raithel/2019,li/2020,lattimer/2010}, but not all of them are able to attain the aforementioned ballpark.

An alternative to elevate the theoretically predicted maximum mass of neutron stars is to consider these objects within the framework of modified theories of gravity. 

Modified theories of gravity are mainly motivated by the crisis observational cosmology is facing through dark energy, dark matter and, more recently, Hubble tension. Dark energy, which is the name given to the mysterious cause of the acceleration in the expansion of the universe, has been investigated through modifications of General Theory of Relativity \cite{wei/2008,wang/2008,bertschinger/2008,nojiri/2005,joyce/2016}. Dark matter, which is the name given to a kind of matter that does not interact electromagnetically, but has a fundamental role in the structure formation in the universe as well as in the galactic dynamics, has also been approached within modified gravity \cite{lisanti/2019,katsuragawa/2017,ferreira/2008,shabani/2023}. The Hubble tension, which is the discrepancy in the measurements of the rate of the expansion of the universe, when comparing local measurements with cosmic microwave background radiation observations, has also been recently treated within modified gravity \cite{sivaram/2022,abadi/2021,mandal/2023,gangopadhyay/2023,adil/2021,montani/2024,odintsov/2021}. 

The applications of modified theories of gravity have been extended to other fields, such as wormholes \cite{pavlovic/2015}, black holes \cite{mann/2022} and stellar structure \cite{olmo/2020,seifert/2007}. As it has been mentioned above, the study of the stellar structure of neutron stars in modified theories of gravity has been considered as a tool for increasing their maximum masses, as one can check \cite{astashenok/2013}.

Nowadays, there is a plethora of alternative gravity theories in the literature. Among those, we quote $f(R)$ gravity \cite{capozziello/2009,capozziello/2005,hu/2007,naf/2011,sotiriou/2010,staykov/2015,papanikolau/2022,song/2007,nojiri/2011,amendola/2007} and $f(R,T)$ gravity \cite{moraes/2019} (check also \cite{harko/2011}), for which $R$ is the Ricci scalar, $T$ is the trace of the energy-momentum tensor, and $f$ is a generic function of the argument, to substitute $R$ in the Einstein-Hilbert action. Another possibility is the $f(R,\mathcal{L})$ theory \cite{harko/2010,azevedo/2016,pourhassan/2020}, for which $\mathcal{L}$ is the matter lagrangian density. As a matter of fact, the $f(R,T)$ and $f(R,\mathcal{L})$ gravity theories were motivated by some flaws, inconsistencies and shortcomings of the $f(R)$ gravity, that can be appreciated, for instance, in \cite{olmo/2007,joras/2011,casado-turrion/2023} (check also \cite{harko/2011}).

As the $f(R,\mathcal{L})$ and $f(R,T)$ gravity formalisms insert both geometrical and material terms in their gravitational action, they are capable of describing a nonminimal coupling between geometry and matter. Below we present some notes on this regard. 

Ludwig et al. have boldly proposed a lagrangian density proportional to $\sqrt{-g}\mathcal{L}^2/R$ and showed it is indistinguishable from Einstein's General Relativity in the dust limit case \cite{ludwig/2015}. It is worth realizing that such a merging of geometry and matter in the same lagrangian satisfies Mach's Principle, since matter cannot exist without curvature and vice-versa. Cosmological models derived from lagrangian densities proportional to $\sqrt{-g}R\mathcal{L}$ and $\sqrt{-g}RT$ were constructed by Gonçalves et al. and Moraes and Sahoo, respectively in \cite{goncalves/2023} and \cite{moraes/2017}. For further insights on the geometry-matter coupling paradigm, one can check \cite{delsate/2012,jimenez/2020}.

Recently, Haghani and Harko have proposed a generalized coupling between geometry and matter, namely the $f(R,\mathcal{L},T)$ gravity \cite{haghani/2021}. The mathematical structure of such a formalism will be given in Section \ref{sec:frlt}. Our main goal in the present paper is to derive the hydrostatic equilibrium configurations of compact astrophysical objects in the $f(R,\mathcal{L},T)$ gravity, which will be made in Section \ref{sec:hecfrlt}, in which we also present our solutions for different equations of state. We present our concluding remarks in Section \ref{sec:dis}.

\section{The $f(R,\mathcal{L},T)$ gravity formalism}\label{sec:frlt}

The $f(R,\mathcal{L},T)$ gravity formalism was recently proposed by Haghani and Harko \cite{haghani/2021} as a possibility for generalizing the geometrical and material terms dependence in the gravitational action. It also allows the generalization of the nonminimal geometry-matter coupling in the action, which for a generalized case, is written as

\begin{equation}\label{action}
    S = \int d^4x \sqrt{-g} \left[\frac{1}{16\pi}\ f(R,\mathcal{L},T) + \mathcal{L}\right],
\end{equation}
in which $f(R,\mathcal{L},T)$ is a function of the argument and we are assuming natural units.

Taking the variation of $S$ with respect to the metric $g^{\mu\nu}$, we obtain

\begin{equation}
    \delta S = \frac{1}{16\pi} \int d^4x \sqrt{-g}
    \left[f_{R}\delta R + \left(f_{T}\frac{\delta T}{\delta g^{\mu\nu}} + f_{\mathcal{L}}\frac{\delta\mathcal{L}}{\delta g^{\mu\nu}}
    - \frac{1}{2}g_{\mu\nu}f - 8\pi T_{\mu\nu} \right) \delta g^{\mu\nu} \right].
\end{equation}
Here, we defined $f_{R}$ = $\partial{f}/\partial{R}$, $f_{T}$ = $\partial{f}/\partial{T}$ and $f_{\mathcal{L}}$ = $\partial{f}/\partial{\mathcal{L}}$. The field equations are obtained from the Palatini identity for the Ricci scalar, namely $\delta R = R_{\mu\nu} \delta g^{\mu\nu} + g_{\mu\nu}\nabla_\rho \nabla^\rho \delta g^{\mu\nu} - \nabla_\mu \nabla_\nu \delta g^{\mu\nu}$, in which $R_{\mu\nu}$ is the Ricci tensor and $\nabla_\alpha$ is the covariant derivative. They are written as

\begin{eqnarray}\label{eq.field}
    &&f_RR_{\mu\nu} - \frac{1}{2}[f(R,\mathcal{L},T) - (f_\mathcal{L} + 2f_T)\mathcal{L}]g_{\mu\nu} + (g_{\mu\nu}\square - \nabla_\mu \nabla_\nu)f_R=\nonumber\\
    &&[8\pi + \frac{1}{2}(f_\mathcal{L} + 2f_T)]T_{\mu\nu} + f_T\tau_{\mu\nu},
\end{eqnarray}
in which the energy-momentum tensor is written as

\begin{equation}
T_{\mu\nu} = - \frac{2}{\sqrt{-g}}\frac{\delta(\sqrt{-g}\mathcal{L})}{\delta g^{\mu\nu}}
\end{equation}
and 

\begin{equation}
\tau_{\mu\nu} \equiv 2g^{\alpha \beta}\frac{\partial^{2}\mathcal{L}}{\partial g^{\mu \nu}\partial g^{\alpha \beta}}.
\end{equation}



To obtain the non-conservation equation for the energy-momentum tensor, we take the covariant derivative of Equation (\ref{eq.field}):

\begin{eqnarray}\label{eq.fieldCovDer}
    &&R_{\mu\nu}\nabla^{\mu}f_R + f_R\nabla^{\mu}R_{\mu\nu}  - \frac{1}{2}\nabla_{\nu}f(R,\mathcal{L},T) =
    \left[8\pi + \frac{1}{2}(f_\mathcal{L} + 2f_T)\right]\nabla^{\mu}T_{\mu\nu} + (\square\nabla_\nu - \nabla_\nu\square)f_R -\nonumber\\
   && \nabla_\nu \left[\left( \frac{1}{2}f_\mathcal{L} + f_T\right) \mathcal{L}\right] + T_{\mu\nu}\nabla^{\mu}\left(\frac{1}{2}f_\mathcal{L} + f_T\right) + \nabla^{\mu}(f_T\tau_{\mu\nu}).
\end{eqnarray}

Using the geometric identity $\nabla^{\mu}G_{\mu\nu} = 0$ and the relation $(\square\nabla_{\nu} - \nabla_{\nu}\square)f_R = R_{\mu\nu}\nabla^{\mu}f_R$, we obtain the equation for the covariant derivative of the energy-momentum tensor as:

\begin{equation}\label{tensoeq}
    \nabla^{\mu}T_{\mu\nu} = \frac{1}{8\pi + f_m}\left[\nabla_{\nu}(\mathcal{L}f_m)
 - T_{\mu\nu}\nabla^{\mu}f_m - \nabla^{\mu}(f_T\tau_{\mu\nu}) - \frac{1}{2}(f_T\nabla_{\nu}T + f_{\mathcal{L}}\nabla_{\nu}\mathcal{L})\right],    
\end{equation}
in which $f_m\equiv f_T + \frac{1}{2}f_{\mathcal{L}}$. In the case of a perfect fluid or scalar field theory, the $\nabla^{\mu}(f_T\tau_{\mu\nu})$ term vanishes.

\section{The hydrostatic equilibrium configurations}\label{sec:hecfrlt}

In the present work, we aim to provide a treatment of neutron stars within the $f(R,\mathcal{L},T)$ theory framework. After deriving the field equations, as well as the equation for the covariant derivative of the energy-momentum tensor, the subsequent step is to choose a functional form for the $f(R,\mathcal{L},T)$ function. In this sense, we follow \cite{haghani/2021} and choose:

\begin{equation}\label{funcform}
    f = R + \alpha \mathcal{L} T,
\end{equation}
in which $\alpha$ is a free parameter of the theory. In our case, the lagrangian density of matter will be set as $\mathcal{L}=-p$, which, together with \eqref{funcform}, reduce the field equations (\ref{eq.field}) to:

\begin{equation}\label{ffes}
R_{\mu\nu}-\frac{1}{2}g_{\mu\nu}R=\left[8\pi-\frac{\alpha}{2}(5p-\epsilon)\right] T_{\mu\nu}-\alpha g_{\mu\nu} p^2.
\end{equation}
To recover General Relativity field equations from (\ref{ffes}), one must simply $\alpha=0$.

\subsection{The TOV-like equation}\label{ss:tov}

The Tolman-Oppenheimer-Volkoff (TOV) equation, firstly introduced in \cite{oppenheimer1939massive, tolman1939static}, describes how the pressure $p(r)$ of a spherically symmetric object changes with respect to the radius $r$ as a function of its enclosed mass $m(r)$ and energy density $\epsilon(r)$. 

To derive the TOV equation for the $f(R,\mathcal{L},T)$ gravity, we choose the adequate static spherically symmetric metric

\begin{equation}\label{metric}
    ds^2=e^{2\psi(r)}dt^2-e^{2\lambda(r)}dr^2-r^2(d\theta^2+\sin^2\theta d\phi^2), 
\end{equation}
with $\psi(r)$ and $\lambda(r)$ being metric potentials.

The $00$ component of the field equations (\ref{ffes}) for metric (\ref{metric}) and the energy-momentum tensor of a perfect fluid, namely

\begin{equation}
    T_\mu^\nu=\texttt{diag}(-\epsilon,p,p,p),
\end{equation}
is

\begin{equation}\label{zerocomp}
  - \frac{e^{-2\lambda}(2\lambda'-1)+1}{r^2}=\left[8\pi-\frac{\alpha}{2}(5p-\epsilon)\right]\epsilon+\alpha p^2,
\end{equation}
for which a prime represents the derivative with respect to the radial coordinate. From this equation we introduce the quantity $m(r)$, to be interpreted as the enclosed mass, such that:

\begin{equation}
    e^{2\lambda}=\left(1-\frac{2m}{r}\right)^{-1},
\end{equation}
and

\begin{equation}\label{massr}
    m(r) = \int^r_0\left[4\pi\epsilon+\alpha(5p\epsilon-\epsilon^2-2p^2)\right]r^2 dr.
\end{equation}

\noindent In addition, in this context, Equation (\ref{tensoeq}) reads:

\begin{equation}\label{consev2}
    p^\prime+\left(p+\epsilon\right)\psi^\prime=-\alpha\left[\frac{ 4p^\prime-e^{2\lambda}\left( \epsilon^\prime-5p^\prime\right) }{16\pi+\alpha\left( \epsilon-5p \right)}\right]p.
\end{equation}

The $11$ component of the field equations is solved to find the following relation:

\begin{equation}\label{psi}
    e^{-2\lambda}(4\psi^\prime r+2)=\left[ 16\pi+\alpha\left(-3p+\epsilon\right)\right]pr^2+2.
\end{equation}

\noindent Then, we can isolate $\psi'$ in (\ref{psi}) to finally write the TOV equation of the $f(R,\mathcal{L},T)$ theory as:

\begin{equation}\label{tov} 
        p'=\frac{-(p+\epsilon)[16\pi+\alpha(-5p+\epsilon)]\{4m+[16\pi+\alpha(-3p+\epsilon)]pr^3\}}{{8r}\left\{8\pi(r-2m)+\alpha\left[m(p-\epsilon)+\frac{1}{2}\left(-p\frac{d\epsilon}{dp}+4p+\epsilon\right)\right]\right\}}.
\end{equation}
By setting $\alpha=0$ the ``usual'' TOV equation is recovered. Note that to derive this equation we defined the energy density in terms of the pressure, $\epsilon=\epsilon(p)$.

\subsection{Equations of state}\label{ss:esos}

To complete the theoretical treatment for neutron stars we must choose an equation of state to close the coupled system of differential equations (\ref{massr}) and (\ref{tov}). This is a relation between pressure and energy density. In possession of it,  we integrate (\ref{massr}) and (\ref{tov}) from the center of the object towards the surface and find the three functions $m$, $\epsilon$ and $p$.

Plenty of equations of state have been proposed to characterize neutron stars, yielding diverse mass-radius relations and varying predictions for their maximum mass. In our analysis, we choose to employ two realistic equations of state sourced from the CompOSE repository \cite{typel2015compose, compose2022compose}. When working with this type of equation of state, it is necessary to interpolate the relation between density and pressure with the tabulated data and then apply to the TOV equation for further integration. The two equations of state we focus on in the present article are the Sly (Skyrme-like) \cite{chabanat1998skyrme} and APR (Akmal-Pandharipande-Ravenhall) \cite{akmal1998equation}.

The Sly equation of state is a comprehensive model used to describe the properties of dense nuclear matter within neutron stars. It is derived from the Skyrme effective nucleon-nucleon interaction, incorporating contributions from both central and spin-orbit terms in the nuclear force. It is parameterized to reproduce experimental data on nuclear matter properties, such as binding energy, saturation density and symmetry energy coefficient. Furthermore, it accounts for nucleon pairing and finite-size effects, offering detailed modelling of neutron star interiors.

In contrast, although also describing the interaction between nucleons, the APR equation of state is formulated based on another theoretical framework, relying on potentials obtained by fitting nucleon-nucleon scattering data. These potentials accurately capture the interactions between nucleons, including both protons and neutrons.

\subsection{Solutions}\label{ss:sol}

Once the equation of state is chosen, the next task involves solving the coupled system comprising Equations (\ref{massr}) and (\ref{tov}), as previously discussed. We then construct a sequence of 200 neutron star models with varying central densities from $2\times 10^{14}~\rm g~cm^{-3}$ to $6\times 10^{15}~\rm g~cm^{-3}$ and also varying $\alpha$ in a range that also encompasses General Relativity outcomes. To ensure consistency, we solve the system by employing two distinct numerical techniques: the Euler method and the 4th-order Runge-Kutta method, both implemented with native packages of Python programming language.

\begin{figure}[h!]
  \includegraphics[width=.5\textwidth]{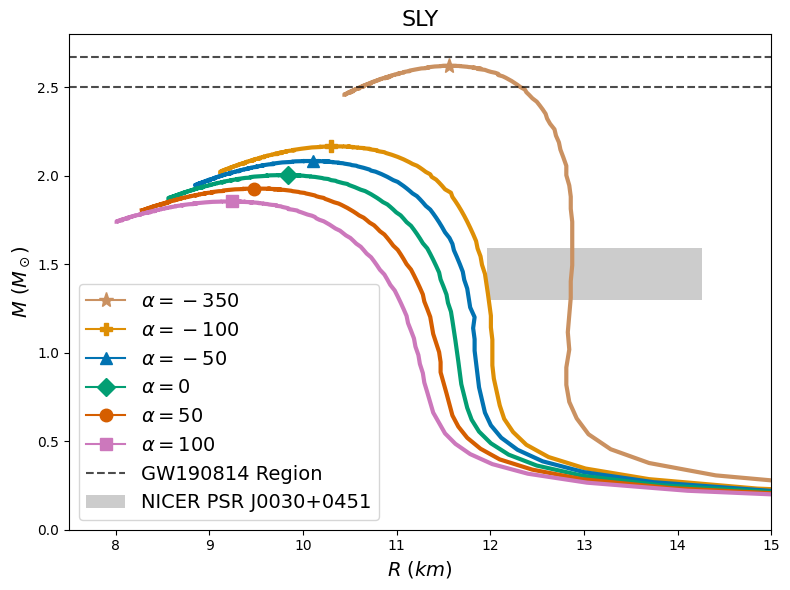}
  \includegraphics[width=.5\textwidth]{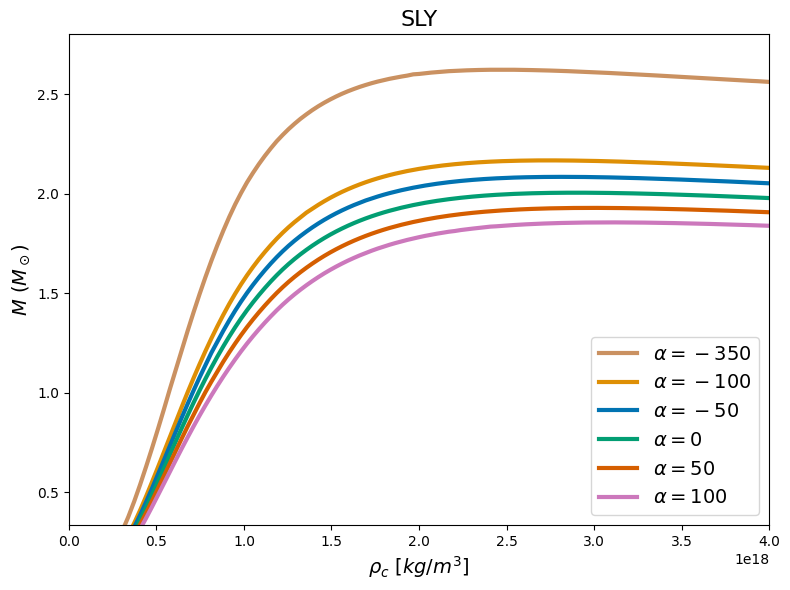}

\caption{Mass-radius (left panel) and mass-central density (right panel) relations for the Sly equation of state considering different values of $\alpha$.}
\label{slymr}
\end{figure}

\begin{figure}[h]
  \includegraphics[width=.5\textwidth]{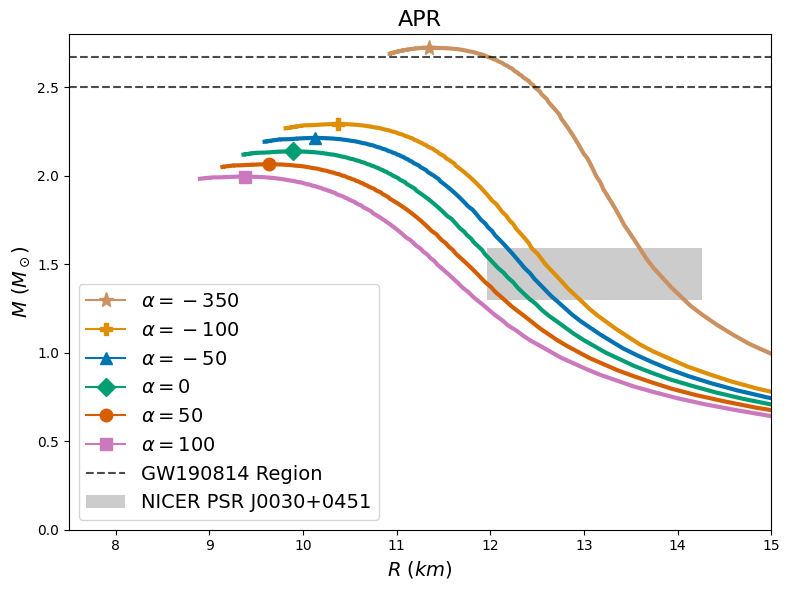}
  \includegraphics[width=.5\textwidth]{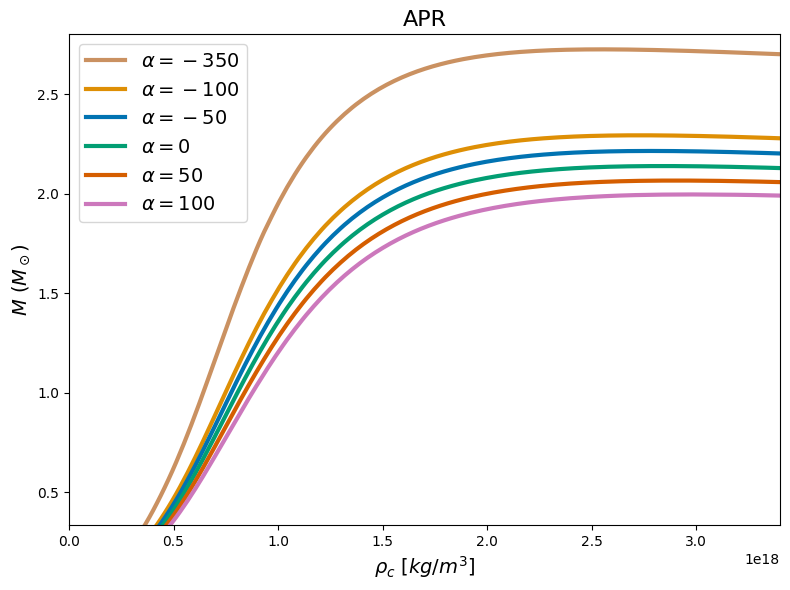}

\caption{Mass-radius (left panel) and mass-central density (right panel) relations for the APR equation of state considering different values of $\alpha$.}
\label{aprmr}
\end{figure}

The results for both Sly and APR equations of state are illustrated in Figures \ref{slymr} and \ref{aprmr}, respectively, and summarized in Table \ref{tabres} within the scope of the investigated $\alpha$ parameter range of the $f(R, \mathcal{L}, T)$ theory. 

\begin{table*}[h!]
\begin{center}
\begin{tabular}[t]{c c c c c}
\hline
\text{EoS} & $\alpha$ & $\rho_c ~ [10^{18}~ km/m^3]$ & $r~[km]$ & $m~[M_{\odot}]$ \\
\hline

\multirow{6}{*}{SLY}
& $-350$ & $2.47$ & $11.57$ & $2.62$\\
& $-100$ & $2.78$ & $10.28$ & $2.17$\\
& $-50$ & $2.80$ & $10.08$ & $2.08$\\
& $0 ~(\text{RG})$ & $2.87$ & $9.81$ & $2.00$\\
& $50$ & $3.02$ & $9.54$ & $1.93$\\
& $100$ & $3.09$ & $9.26$ & $1.85$\\
\hline
\multirow{6}{*}{APR}
& $-350$ & $2.54$ & $11.34$ & $2.73$\\
& $-100$ & $2.73$ & $10.37$ & $2.28$\\
 & $-50$ & $2.77$ & $10.13$ & $2.20$\\
& $0 ~(\text{RG})$ & $2.84$ & $9.89$ & $2.14$\\
& $50$ & $2.91$ & $9.62$ & $2.06$\\
& $100$ & $2.95$ & $9.38$ & $1.99$\\
\hline

\hline
\end{tabular}
\caption{The summarized findings regarding the two equations of state examined in this work for the considered range of the coupling parameter $\alpha$.}
\label{tabres}
\end{center}
\end{table*}

Currently, the precision of observational measurements is important in validating or refuting modified theories of gravity and the equations of state adopted for the description of neutron stars. In this context, we consider the region defined by the statistical confidence interval associated with the $\rm GW190814$ gravitational wave event \cite{abbott/2020}. This event is related to the merger of two compact objects, with the secondary one presumed to be a neutron star with a mass of $2.59^{+0.08}_{-0.09}~\rm M_\odot$. Another constraint that arises directly from observational data is that from the millisecond pulsar PSR J0030+0451, as observed by the Neutron Star Interior Composition Explorer (NICER) collaboration \cite{miller2019psr}. From X-ray measurements, the team estimated the radius and the mass of the pulsar to be $13.02^{+1.24}_{-1.06}~\rm km$ and $1.44^{+0.15}_{-0.14}~\rm M_{\odot}$, respectively. We picture this constraint as a grey rectangle in Figures \ref{slymr} and \ref{aprmr} below, in their left panel. In their right panels, we plot the total mass of the neutron star as a function of the central density. Different curves mean different values of $\alpha$.

In Fig.1 left panel we show that the results concerning the Sly equation of state reveal that within General Relativity framework ($\alpha=0$), both the maximum mass and radius fall outside the specified regions of interest, indicating an inability to accurately represent observed data using this equation of state alone. Similarly, decreasing the value of $\alpha$ yields comparable outcomes. However, our plots show that when opting for larger values of $\alpha$, the curve begins to converge toward the constrained regions, eventually intersecting both constraints.

In contrast, according to Figure \ref{aprmr} left panel, regarding the results for the APR equation of state within Einstein's General Theory of Relativity theoretical framework, the curve lies inside the boundary outlined by NICER data. However, it fails to meet the strong constraint imposed by the maximum mass derived from the $\rm GW190814$ gravitational wave event. When increasing the value of the parameter $\alpha$, the curve surpasses the maximum mass delineated by the dashed lines. This shows that by using the APR equation of state in the context of $f(R,\mathcal{L},T)$ gravity theory, it is possible to obtain even higher maximum masses than those presently detected.

The behavior for both equations of state studied in this paper aligns with expectations. As indicated in Table \ref{tabres}, higher central densities correspond to smaller radii for neutron stars. In this context, while not a definitive criterion for verifying neutron star stability, an useful metric is the rate of change of mass with respect to radius. If this rate is greater than zero, $dm/dr>0$, the star cannot achieve stable equilibrium. By inspecting the mass-radius curves depicted in both Figures \ref{slymr} and \ref{aprmr}, it becomes evident that the solutions on the left side of the maximum mass are unstable.

\section{Discussions and concluding remarks}\label{sec:dis}

As the observational data on neutron stars is increasing, so is the challenge in modelling such objects. On this regard, the extended gravity formalism rises as an optimistic approach. In fact, neutron stars provide a ``laboratory'' for testing gravity in strong regimes.

In the present paper we have derived the TOV-like equation for the particular $f(R,\mathcal{L},T)$ gravity formalism, recently proposed by Haghani and Harko. This theory is capable of recovering not only Einstein's General Relativity, but also $f(R)$, $f(R,\mathcal{L})$ and $f(R,T)$ gravity formalisms, in particular cases. In this way, it is referred to as a ``generalized geometry-matter coupling theory of gravity''.

Here, we have solved the TOV-like equation for two realistic equations of state, namely Sly and APR equations of state. We have tested our models with observational data coming from LIGO/Virgo and NICER teams. 

Remarkably, for the Sly equation of state, General Relativity is uncapable of according with observational data. We have shown that for this equation of state, it is possible to have neutron star models that adjust with data only for higher contributions from the extended gravity theory. Otherwise, using the APR equation of state, the observational data coming from NICER is attained for a great range of values of $\alpha$, while data coming from LIGO/Virgo regarding a massive neutron star is only attained for higher values of $\alpha$, i.e., higher contributions from extended gravity.

In closing, the current article has the potential for expansion across various fronts and we briefly introduce some of them below. Different functional forms can be assumed for the $f(R,\mathcal{L},T)$ function. Those will invariably imply in different TOV-like equations to be solved. Different equations of state can be used to describe matter inside neutron stars. The TOV-like equation here presented can be used to generate hydrostatic equilibrium configurations of other compact objects, such as white dwarfs. On this regard, the importance of relativistic effects in the structure of massive white dwarfs is outlined in \cite{carvalho/2018}. Finally, a further stability test for the neutron stars here obtained could be the test of stability against radial pulsations.

\acknowledgments

J.A.S. Fortunato thanks FAPES for financial support. JG de Lima Júnior thanks CAPES for financial support.


\bibliographystyle{JHEP}
\bibliography{Bib.bib}



\end{document}